\documentclass[12pt]{amsart}
\usepackage{enumerate}
\usepackage{enumitem}
\usepackage[colorlinks=true, linkcolor=blue, urlcolor=blue, citecolor=blue, anchorcolor=blue, pdfborder={0 0 0}]{hyperref}
\usepackage{url}
\usepackage{breqn}
\usepackage{graphicx,color,colortbl}
\usepackage[dvipsnames,table,xcdraw]{xcolor}
\usepackage{cite}
\usepackage{amsthm, amsmath, amssymb}
\usepackage{mathtools}
\usepackage[top=45truemm, bottom=45truemm, left=30truemm, right=30truemm]{geometry}
\usepackage{nicefrac}
\usepackage{cancel}
\usepackage{float}
\usepackage{tabularx}
\usepackage{makecell}
\usepackage{array}
\usepackage{ragged2e}
\usepackage{graphicx}
\usepackage{caption}
\usepackage{subcaption}
\usepackage[symbol]{footmisc}
\usepackage{tikz}
\makeatletter
\def\@settitle{\begin{center}%
  \baselineskip14\p@\relax
  \bfseries
  \uppercasenonmath\@title
  \@title
  \ifx\@subtitle\@empty\else
     \\[1ex]\uppercasenonmath\@subtitle
     \footnotesize\mdseries\@subtitle
  \fi
  \end{center}%
}
\def\subtitle#1{\gdef\@subtitle{#1}}
\def\@subtitle{}
\makeatother

\definecolor{bg}{rgb}{0.95,0.95,0.95}
\definecolor{LightGray}{RGB}{242,242,242}

\newcommand{\code}[1]{\colorbox{bg}{\texttt{#1}}}

\newcommand{\ket}[1]{\vert {#1} \rangle}

\newcommand{\bra}[1]{\langle {#1} \vert}

\newcolumntype{P}[1]{>{\RaggedRight\hspace{0pt}}p{#1}}
\newcolumntype{L}[1]{>{\raggedright\let\newline\\\arraybackslash\hspace{0pt}}m{#1}}
\newcolumntype{C}[1]{>{\centering\let\newline\\\arraybackslash\hspace{0pt}}m{#1}}
\newcolumntype{R}[1]{>{\raggedleft\let\newline\\\arraybackslash\hspace{0pt}}m{#1}}

\theoremstyle{definition}

\title{Predict better with less training data using a QNN}

\author[Barry Reese]{Barry D. Reese}
\address{Barry D. Reese\\Capgemini\\Olaf-Palme-Straße 14\\Munich, Germany}
\curraddr{}
\email{barry.d.reese@gmail.com}

\author[Marek Kowalik]{Marek Kowalik}
\address{Marek Kowalik\\Capgemini\\Legnicka 48H, 54-202\\ Wrocław, Poland}
\curraddr{}
\email{marek-jozef.kowalik@capgemini.com}

\author[Christian Metzl]{Christian Metzl}
\address{Christian Metzl\\Capgemini\\Olaf-Palme-Straße 14\\Munich, Germany}
\curraddr{}
\email{christian.metzl@capgemini.com}

\author[Christian Bauckhage]{\href{https://orcid.org/0000-0001-6615-2128}{\includegraphics[scale=0.06]{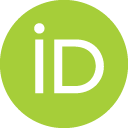}}\hspace{1mm}Christian Bauckhage}
\address{Christian Bauckhage\\Fraunhofer IAIS\\IZ Birlinghoven\\St.~Augustin, Germany}
\curraddr{}
\email{christian.bauckhage@iais.fraunhofer.de}

\author[Eldar Sultanow]{\href{https://orcid.org/0000-0001-5257-2236}{\includegraphics[scale=0.06]{orcid.png}}\hspace{1mm}Eldar Sultanow}
\address{Eldar Sultanow\\Capgemini\\Bahnhofstraße 30\\Nuremberg, Germany}
\curraddr{}
\email{eldar.sultanow@capgemini.com}


\begin{document}

\begingroup
\let\MakeUppercase\relax
\clearpage\maketitle
\thispagestyle{empty}
\endgroup

\begin{abstract}
Over the past decade, machine learning revolutionized vision-based quality assessment for which convolutional neural networks (CNNs) have now become the standard. In this paper, we consider a potential next step in this development and describe a quanvolutional neural network (QNN) algorithm that efficiently maps classical image data to quantum states and allows for reliable image analysis. We practically demonstrate how to leverage quantum devices in computer vision and how to introduce quantum convolutions into classical CNNs. 
Dealing with a real world use case in industrial quality control, we implement our  hybrid QNN model within the PennyLane framework and empirically observe it to achieve better predictions using much fewer training data than classical CNNs. In other words, we empirically observe a genuine quantum advantage for an industrial application where the advantage is due to superior data encoding. 
\end{abstract}

\section{Introduction}
\label{sec:introduction}

In this paper, we are concerned with quantum computing for computer vision and the obvious immediate question is: why? After all, recent progress in design and training of (variants of) convolutional neural networks (CNNs) has led to reliable vision systems for a wide range of practical applications. However, in order to perform well, modern neural networks typically need training with large amounts of representative data. This limits their applicability in settings where such data are not readily available, for instance, in quality control of manufactured parts where images of defective examples are rare. It is in contexts like these where quantum computing could play a role.

While issues pertaining to insufficient training data prompted the applied machine learning community to investigate  hybrid learning systems which combine data- and knowledge-driven techniques to obtain reduced-parameter models for representation learning in small data settings  \cite{Oyallon2019-SNF,Karniadakis2021-PIM,vRueden2021-IML}, recent work on quantum machine learning suggests that quantum encodings involving superposition and entanglement can provide expressive representations, too \cite{Dunjko2016quantum}. Given the predominance of CNNs in modern image analysis, it is therefore no surprise that quantum analogs of neural networks attract increasing attention. For instance, Kerenidis et al.~\cite{Kerenidis_2019} are interested in quantum deep learning and develop a quantum algorithm for accelerating  use and training of deep CNNs. In numerical simulations, they classify MNIST images and demonstrate their approach to work. However, their algorithm crucially hinges on the assumption of the availability of quantum random access memories (QRAMs) which, as of this writing, are not yet practically feasible. 

Yet, technically feasible methods do exists as well. For instance, Lü et al. \cite{Lue_2021} suggest \emph{quantum convolutional neural network} (QCNN). They describe deployable quantum circuits whose architecture is analogous to that of classical CNNs and perform numerical simulations with MNIST data which demonstrate the effectiveness of their model. Henderson et al.\cite{Henderson2021-QNN,Henderson2021-MFA} note that the benefits of CNNs for image analysis largely stem from their ability to learn hierarchical feature representations of the given data. As such features are extracted using several convolutional layers, they consider quantum convolutional layers or \emph{quanvolutional} layers for short. That is, they transform input data bey means of several random quantum circuits and use the resulting representations for classification. Experiments with MNIST and satellite image data  show that their \emph{quanvolutional neural networks} (QNNs) have higher test accuracy and faster training times than comparable classical CNNs. 

Similar approaches have been pursued by Mathur et al.~\cite{Mathur2021-MIC} and Mattern et al.~\cite{Mattern_2021}. The former propose quantum orthogonal neural networks and benchmark them on medical image data which reveals limitations of present day quantum hardware. The latter work with QNNs in a variational- or quantum-classical hybrid setting which allows them to train quanvolutional layers. They experiment with MNIST data, evaluate different quantum image encoding schemes, and find that there does not seem to be a single best encoding choice. Finally, Cong, Choi, and Lukin \cite{Cong_etal_2019} also work with variational QNNs and improve their efficiency w.r.t. the number of required variational parameters but do not experiment with image data. 

Our approach in this paper, too, relies on QNNs as proposed in \cite{Henderson2021-QNN} and integrates them into a classical CNN stack similar to the manner in \cite{Cong_etal_2019}. Our main focus is on applying QNNs to a genuine practical use case for intelligent automation in industry. Specifically, we integrate quantum convolution into a classical CNN for vision-based quality control. The data we are working with consists of top view images of manufactured parts and the task is to automatically detect deficient parts, i.e. parts with cracks.

Our main and noteworthy empirical result is the following: In the scenario considered, our QNN model requires less training data to achieve better predictions than classical CNNs. In other words, we observe a genuine quantum advantage for a real world application scenario. This advantage is not so much due to quantum speedup but rather due to superior quantum encoding of small data sets. Put differently, the observed quantum advantage is not preconditioned on the capabilities of still mainly theoretical universal quantum computers but can already be harnessed on present day quantum devices. 

Our implementation of the hybrid QNN model for quality control is divided into two development stages. First, we use quantum convolution to map classical input image data into quantum state spaces. Second, we extend and optimize that basic procedure. Especially for the model from stage 2, we experimentally find  higher prediction accuracy. To illustrate this observation, we perform extensive evaluations in which we compare the performances of classical CNNs and QNNs from stages 1 and 2. 

The stage 1 is to understand the impact of quantum circuit encoding image data on the number of data inputs required to train a QNN  and compared the performance against a benchmark CNN.  Stage 2 improves the stage 1 algorithm to understand how quantum technologies might be used in industrial applications and the effect on the number of data inputs.  Our purpose is not to build the best CNN we can and compare it against the best QCNN we can build.  Rather, we intend to gain an understanding of how the quantum circuit affects data representations and their impact on training computer vision models. Our procedures are as follows: 

Stage 1:  Build a basic 4 qubit quanvolutional neural network and train with 100 and 50 training inputs and compare against classical benchmark.
\begin{itemize}
  \item Pre-process image data with 4 qubit quantum circuit and convert to quantum tensor.
  \item Ingest quantum tensor 100 train inputs to classical CNN for the QNN model.
  \item Invest same 100 classical training images to similar CNN as benchmark.
  \item Reduce the number of training inputs by half for both models to see how they perform with fewer inputs.
  \item Compare the results.
\end{itemize}
Stage 2:  Scale up Stage 1 to 16 qubits and a more sophisticated pre-processing algorithm to process high resolution images with quantum circuits.  Train with a 50/50 train/test split and compared against a 40/60 train test split.
\begin{itemize}
  \item Integrate an image splitting and weighting algorithm in order to process high contrast images with quantum circuits.  Splitting also serves to localize surface cracks in images.
  \item Pre-process image data with a 16 qubit quantum circuit and convert to a quantum tensor.
  \item Ingest the quantum tensor with 50/50 train/ test split to classical CNN for the QNN model.
  \item Use the same 50/50 train/split classical training images for a similar CNN as a benchmark.
  \item Cut number of training inputs to 40/60 for both models to see how the models perform with fewer inputs.
  \item Compare the results.
\end{itemize}

Our presentation proceeds as follows: First, we ever so briefly recall the mathematical foundations of quantum computing and discuss challenges regarding implementations on present day quantum hardware. Then we present details as to the implementations of our QNN models and results obtained in practical experiments. We then discuss these results and finally conclude with a summary of our main results and an outlook to auspicious future work.

\section{Quantum Computing}

In order for this paper to be self-contained, we next briefly summarize the mathematical foundations of quantum computing and discuss practical challenges w.r.t.~current quantum hardware. readers familiar with these topics may safely skip ahead.

In quantum computing, the equivalent to classical bits are quantum bits or qubits which are two-state quantum mechanical systems. Mathematically, the state of such a system is modeled as a vector in the Hilbert space $\mathbb{C}^2$ and, written in Dirac notation, a commonly considered orthonormal basis of this space is given by the vectors $\ket{0}$ and $\ket{1}$. This notation alludes to the two possible states $0$ and $1$ of a classical bit, however and crucially, a qubit $\ket{q}$ can exist in a superposition $\ket{q} = a_0 \ket{0} + a_1 \ket{1}$ where the coefficient $a_0, a_1 \in \mathbb{C}$ obey the normalization condition $\lvert a_0 \rvert^2 + \lvert a_1 \rvert^2 = 1$. Although a qubit can thus exists in infinitely many different state, we can still only extract one classical bit of information from it. This is due to the quantum mechanical phenomenon of (wave function) collapse upon measurement. Specifically, if we perform a measurement on $\ket{q} = a_0 \ket{0} + a_1 \ket{1}$, it will probabilistically collapse to either of its basis states $\ket{0}$ or $\ket{1}$. The probability that the resulting state is $\ket{0}$ is given by the amplitude $\lvert a_0 \rvert^2$ and the probability for it being $\ket{1}$ is given by the amplitude $\lvert a_1 \rvert^2$.

A system of $n$ qubits $\ket{q_1}, \ket{q_2}, \ldots, \ket{q_n}$ exists in the Hilbert space $\mathbb{C}^2 \otimes \mathbb{C}^2 \otimes \ldots \otimes \mathbb{C}^2$ and can thus be in a superposition of $2^n$ basis states $\ket{00\ldots0}, \ket{00\ldots1}, \ldots, \ket{11\ldots1}$. It is this exponential growth of state spaces in the number of constituent qubits that provides an advantage of quantum computing over classical computing. Moreover, there exist states which cannot be described in terms of their constituent qubits separately. Such states do not have classical analogues and are called entangled states. Entanglement allows for ideas such as super dense coding and thus provides another quantum advantage. 

The temporal dynamics of an unmeasured qubit systems is governed by the Schr\"odinger equations. Since any evolution of the system must keep the normalization condition intact, operations on the system must necessarily by unitary. That is, quantum operators $U : \mathbb{C}^{2 \otimes_n} \rightarrow \mathbb{C}^{2 \otimes_n}$ obey $U U^\dagger = U^\dagger U = I$ where $\dagger$ denotes conjugate transposition. In the quantum gate approach to quantum computing, quantum algorithm developers are thus concerned with quantum circuits composed of unitary operators which act (in sequence or in parallel) on $n$ qubits so as to transform an input state to a desired output state.


Finally, we observe that quantum algorithm design typically considers properties of logical qubits rather than of physical qubits. The former are a mathematical idealization of the latter which are physical systems realized in the hardware of a quantum computer. It is therefore important to note that currently existing quantum computers are so called noisy intermediate scale quantum (NISQ) devices \cite{Preskill_2018}. Such machines are still limited w.r.t.~the number of physical qubits they can manipulate and thus with the number of logical qubits they can realize. They also come with limitations as to accuracy of quantum gates and their overall fault tolerance. All of this means that, in the NISQ era of quantum computing is era, real quantum advantages are still hard to come by and it is crucial to find use cases which are easy for existing quantum machines and hard for classical ones. This is because, in order to scale up quantum algorithms, one needs to encode information in highly entangled states. To do this free of error, one must protect the state  from outside interference which could lead to decoherence. 

The practical use case we consider in this paper requires the classification of entire images. From the point of view of currently feasible quantum computing, this poses a significant challenge. Potential approaches to mitigate potential loss of information include QGANs to resize images or quanvolutional layers over local subsections of the data, building circuits with more qubits, and changing the architecture of the neural networks themselves. For the practical examples discussed below, we opted for adding quanvolutional layers over local subsections of the data. Practical computations were performed using a quantum simulator running on an Amazon Braket machine with \href{https://pennylane.ai/}{PennyLane.AI} to interact with the hardware and \href{https://keras.io/}{TensorFlow Keras} as a neural network training framework.

\section{Quanvolutional Neural Network}
\label{sec:quanvolutional}
In this section We describe the fashion our quantum embedding has been created. Basically, our approach relies on the idea of a quanvolutional neural network \cite{Henderson2021-QNN} and utilizes quantum devices for computer vision purpose via quantum convolution in a classical CNN stack similarly as described in reference \cite{Cong_etal_2019}.

\subsection{Stage 1}
\begin{figure}[t!]
	\centering
	\includegraphics[clip, trim=0cm 0cm 0cm 0cm, width=1\textwidth]{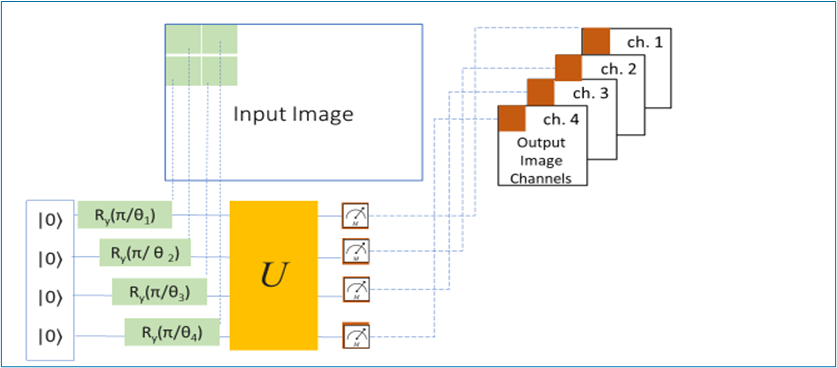}
	\caption{Mapping classical data to a quantum space for better representation.}
	\label{fig:1}
\end{figure}

\begin{figure}[t!]
	\centering
	\includegraphics[clip, trim=0cm 0cm 0cm 0cm, width=0.6\textwidth]{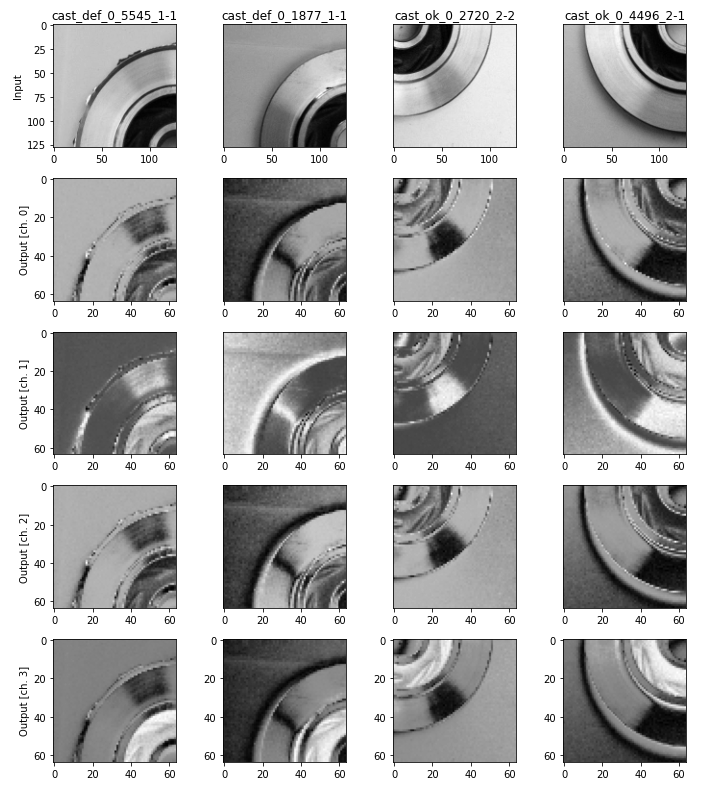}
	\caption{Effect of mapping classical data to quantum space using a quantum convolution.}
	\label{fig:3}
\end{figure}

In the first stage, we observed the impact of a 4 qubit Quantum Circuit with regards to the data inputs required to train a QNN versus a Classical CNN. First we convolved the input image with many applications of random quantum circuits on input u spatially local $n\times n$ kernels (see Figure~\ref{fig:1}). Second, our measurement consisted of a Pauli-Z gate which is $Z=\ket{0}\bra{0}-\ket{1}\bra{1}$ to produce our quantum encoding where output $o_x$ is the quantum state $q(i_x)$ (see Figure~\ref{fig:3}).

\begin{figure}[t!]
	\centering
	\includegraphics[clip, trim=0cm 0cm 6cm 1cm, width=0.8\textwidth]{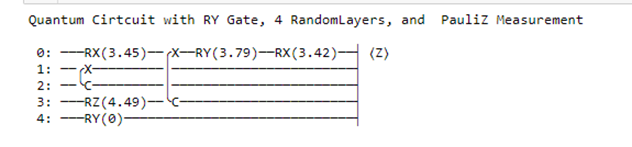}
	\caption{Quantum circuit with RY gate, $4$ random layers, and Pauli-Z measurement.}
	\label{fig:2}
\end{figure}

In Figure~\ref{fig:1}, $\ket{0}$ is the orthogonal $z$-basis state of a qubit; by default, all qubits are initialized to $\ket{0}$ state. The operator $R_y(\pi/\theta_n)$ performs a single qubit rotation through angle $\theta$ around the $y$ axis scaled by a factor of $\pi$. The process for quantum encoding takes is denoted by the unitary operator $U$. It is the analog to the convolutional filters in a classical network. The corresponding quantum circuit is shown in Figure~\ref{fig:2}.

For the model architecture, we chose the Keras sequential model. This is appropriate when modeling a plain stack of layers where each layer has one input tensor and one output tensor. Indeed, this is appropriate for both the CNN and QNN where the CNN takes a classical $32$-bit input and the QNN takes the local quantum embeddings as inputs. 

Using this architecture, we found the model where the QNN and CNN performed equally on the 100 training inputs. Then we reduced the number of training inputs to 50 to investigate how the models would perform with less training data.

We evaluated the results by analyzing how well the models fit to the training data and made sure that loss was decreasing consistently across epochs. Then we evaluated how well the models performed on unseen data and analyzed how well the loss decreased consistently across epochs.

\begin{figure}[t!]
	\centering
	\includegraphics[clip, trim=0cm 0cm 0cm 0cm, width=0.9\textwidth]{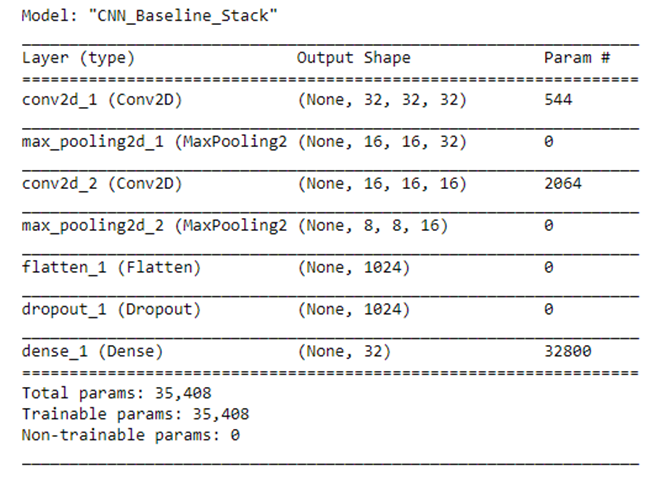}
	\caption{Baseline CNN architecture from the KERAS summary.}
	\label{fig:4}
\end{figure}

For our CNN model, we added an additional conv2d layer to get an even comparison with our QNN model. Our model architecture looks like this for the CNN Model. Notice that the input shape is $n\times32\times32\times32$. This represents a shape of (batch size, height, weight, depth). The depth of $32$ represents the number of \code{conv2d} filters applied in this layer. Figure~\ref{fig:4} illustrates our CNN Model Architecture. 

\begin{figure}[t!]
	\centering
	\includegraphics[clip, trim=0cm 0cm 0cm 0cm, width=0.9\textwidth]{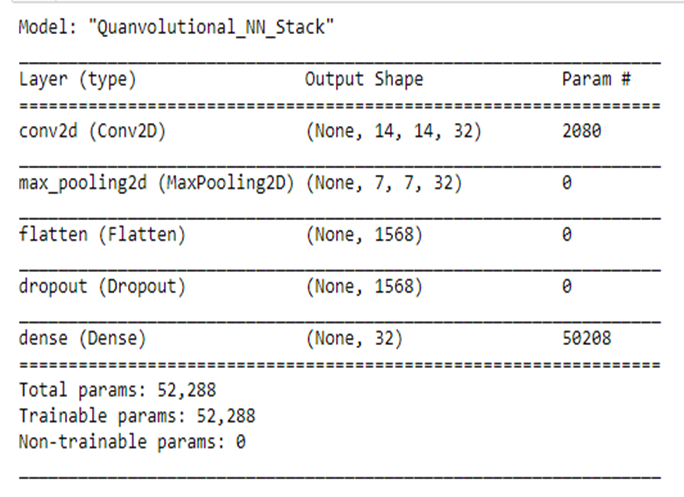}
	\caption{QCNN architecture from the KERAS summary.}
	\label{fig:5}
\end{figure}

Since we do the quanvolutional embedding in our QNN, we take out one of the \code{conv2d} layers and one \code{max_pooling2d} layer from our model. Figure~\ref{fig:5} illustrates our QNN Model Architecture.

The CNN model starts with $32\times32$ image size and $32$ filters in its convolutional layer with a kernel size of $4$, and pools this by factor of 2 in its second layer. On the other hand, our QNN model starts with an input size of 14 with the same kernel size. This does not strictly give an equal model because the CNN input size is larger, and thus has an advantage. We can make up for this advantage in our further work by performing our quantum operation over the $32$-bit image rather than on a subset of the image.

\begin{figure}[t!]
	\centering
	\includegraphics[clip, trim=0cm 0cm 0cm 0cm, width=0.9\textwidth]{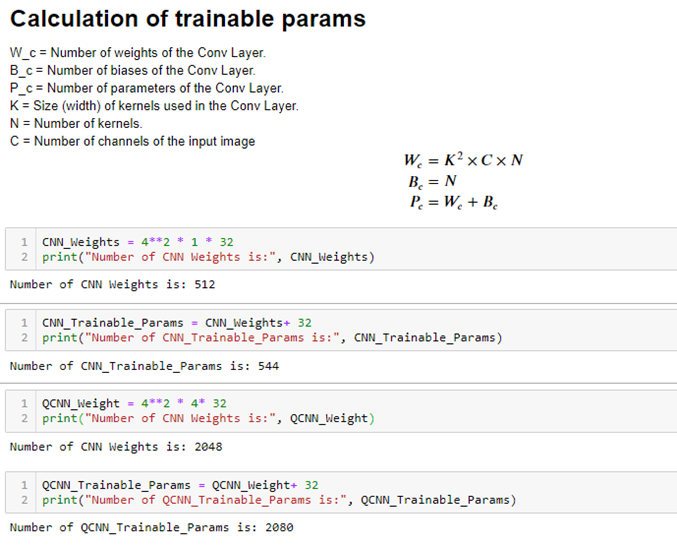}
	\caption{Computation of trainable parameters in the accompanying Jupyter notebook.}
	\label{fig:6}
\end{figure}

For our fully connected layer, we gave each model $32$ fully connected units. Since the QNN fully connected input was denser from the quantum embedding we had $1568$ hidden units, while the CNN had $1024$ hidden units, and each had 32 fully connected units and a dropout of 0.  Normally we would make our fully connected units equal to the number of classes we are trying to classify. However, $32$ was the level at which model performance was the most equal. For this set of experiments, we will leave it at $32$ and change it in the next steps to get the most accurate QNN possible.
We will include a brief discussion of trainable parameters because one might think it looks strange when you see a smaller input tensor with a larger number of trainable parameters. The main reason for this is Pauli-Z measurement which produced the $4$ classical output values, thus $4$ channels. The effect is that we have a denser input tensor for the QNN model. However, since we were comparing the QNN to the CNN and not the other way around, we left a few parameters the same. For example, the multiplier for the QNN weight should have been $14$ rather than $32$. Also, the number of biases in the convolutional layer should have been 14 rather than $32$. Tuning these parameters properly will most likely give a better model in our next steps. The computation for the fully connected layer and total parameters from the architecture stack builds off this. Therefore, it is not necessary to draw the full calculation. Figure~\ref{fig:6} shows the computation of trainable parameters.

We compiled our model with the Adam optimizer \cite{KingmaBa}, sparse categorical cross-entropy for the loss and optimized the used the accuracy as our loss metric. Adam is a popular optimizer because it can create adaptive estimates during training and is thus easier to tune and more computationally efficient. It is also known to better at correcting for bias than other optimizers. The sparse categorical cross-entropy is appropriate when there are two or more label classes. In our case we have 2: cracked vs. non-cracked. It takes an integer indicating the classes (1 or 0 in our case). Finally, this loss is appropriate when we want to calculate the most likely matching category rather than a probably match. There are other loss functions we could have used, but sparse categorical cross-entropy gives us a good baseline. In other QNN implementations, custom loss metrics are used.

\subsection{Stage 2}
\label{sec:Stage_2}
In the second stage, we observe how our our algorithm performs with more realistic image inputs one might find in an industrial application. Since our algorithm affects the earlier convolutions heavily, we think that the quantum embeddings strongly assist in helping the model built the lower-level abstractions from the data. As a result, we have achieved a much higher level of success, namely a much higher prediction accuracy over stage 1. The key differences are:

\begin{itemize}
    \item Splitting of the images for localization.
    \item Using the masks to label the images.
    \item Stitching the images back together after model execution for interpretability.
    \item Use of a 16-qubit simulator instead of a 4-qubit simulator in Stage 1.
\end{itemize}

Our algorithmic approach in stage 2 includes the following steps:

\begin{enumerate}
    \item We begin by loading the data from local directories.
    \item We split the images into 9 x 9 individual images for defect localization. Due to NISQ simulators, we cannot perform quantum preprocessing of large images without significant data loss. Furthermore, due to the large resolution size of these images and the relatively small localization of the cracks, these cracks are very hard to detect in the classical way.
    \begin{itemize}
        \item In order to achieve minimal loss from re-scaling the images, we developed an algorithm to split them into regions.
        \item The region is 9 x 9 of the original.
        \item 9 represents the common denominator between all of the image sizes in the data. There were 3 different image sizes.
        \item For each image size, we assigned a padding and re-scaled to get to a factor of 9.
        \item We then used the resized images to assign binary (positive, negative) and region labels to each region of the image.
        \item We then performed pre-processing and predictive modeling of the regions.
        \item We then stitched the images back together with the localization and the label.
    \end{itemize}
    \item We perform quantum embeddings on the data including an Ry gate and a Pauli-Z measurement.
    \item For our experiment, we construct two models.
    \begin{itemize}
        \item One is a QNN taking the quantum embedded input tensor of the regions and labels as input.
        \item The other is a classical CNN taking the classical input tensor of the regions and labels as input.
        \item We find the model architecture that does an even comparison between the CNN and QNN.
        \item We analyze the results.
    \end{itemize}
\end{enumerate}

\begin{figure}[t!]
	\centering
	\includegraphics[clip, trim=0cm 0cm 0cm 0cm, width=0.8\textwidth]{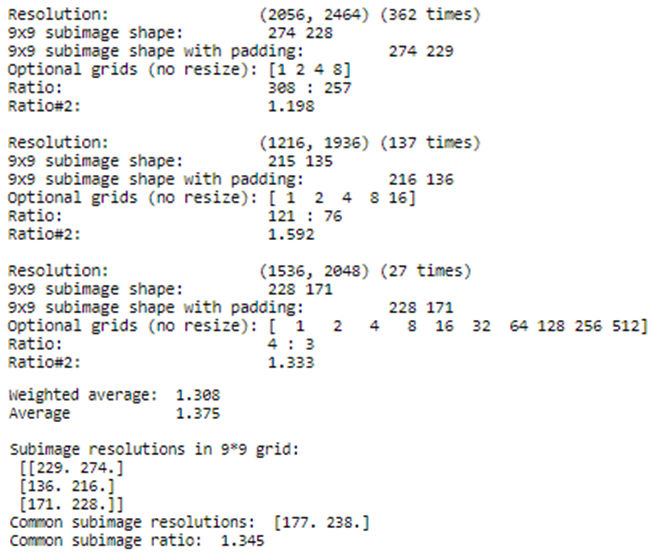}
	\caption{Computation of our $9\times9$ grid image weights.}
	\label{fig:7}
\end{figure}

Coming back to the aspect of resizing and padding our images in $9\times9$ grid, we note that three different categories of resolutions are present in the given images of stage 2. Therefore, we weigh the created $9\times9$ grids differently for each resolution category to achieve a common factor of 9 while having to pad the edges of the split images in order to account for split differences. Figure~\ref{fig:7} shows those resizing ratios as weights.

One can extend the same idea also to the context of quantum variational circuits (VQCs) for transfer learning, but again we opine on the work of Henderson et al.~\cite{Henderson2021-QNN} and proceed with the QNN approach:

\begin{enumerate}
    \item A small region of the input image, in our example a 4 × 4 square, is embedded into a quantum circuit. In this demo, this is achieved with parameterized rotations applied to the qubits initialized in the ground state.
    \item A quantum computation, associated to a unitary U, is performed on the system. The unitary could be generated by a variational quantum circuit or, more simply, by a random circuit as proposed by Henderson et al.~\cite{Henderson2021-QNN}. We will use a random circuit with 4 random layers in our approach.
    \item The quantum system is finally measured, obtaining a list of classical expectation values. The measurement results could also be classically post-processed as proposed but, for simplicity, in this demo we directly use the raw expectation values.
    \item Analogously to a classical convolution layer, each expectation value is mapped to a different channel of a single output pixel.
    \item Iterating the same procedure over different regions, one can scan the full input image, producing an output object which will be structured as a multi-channel image.
    \item The quantum convolution is only followed by classical layers.
\end{enumerate}

\begin{figure}[t!]
	\centering
	\includegraphics[clip, trim=0cm 0cm 0cm 0cm, width=1\textwidth]{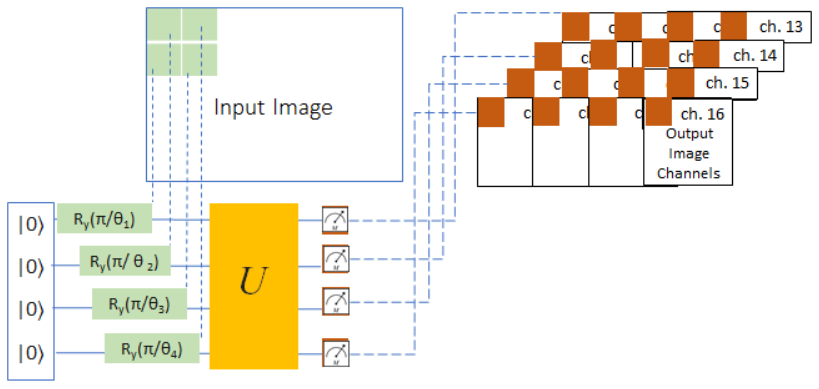}
	\caption{Quantum embedding with $16$ output channels representing $16$ qubits.}
	\label{fig:8}
\end{figure}

The main difference with respect to a classical convolution is that a quantum circuit can generate highly complex kernels whose computation could be classically intractable. In Stage 1 in this paper, we used $4$ qubits in our quantum simulator. Then in Stage 2 phase we scaled it up to $16$ qubits. Figure~\ref{fig:8} depicts our quantum embedding whereby four basis state qubits are mapped via a unitary transformation from an input image to $16$ Output Image Channels.

The notation is consistent to the one used for Figure~\ref{fig:1}, see also Glossary in Appendix~\ref{appx:glossary}. The unitary operation is $4$ times larger to bring it from $4$ qubits to $16$ qubits, but the operation is too large to be shown in one chart. There actually be $16$ gates, but for simplification the above chart only shows $4$ of them.

The model circuit and the embedding effects on the images in Stage 2 are extremely lengthy. If you would like to see the circuit drawing, and the embedding effect on images, please see the attached Jupyter notebook. They are mostly the same, the effect of the padding in the images is apparent. Furthermore, the quantum circuit is larger by a factor of $4$.

\begin{figure}[t!]
	\centering
	\includegraphics[clip, trim=0cm 0cm 0cm 0cm, width=0.8\textwidth]{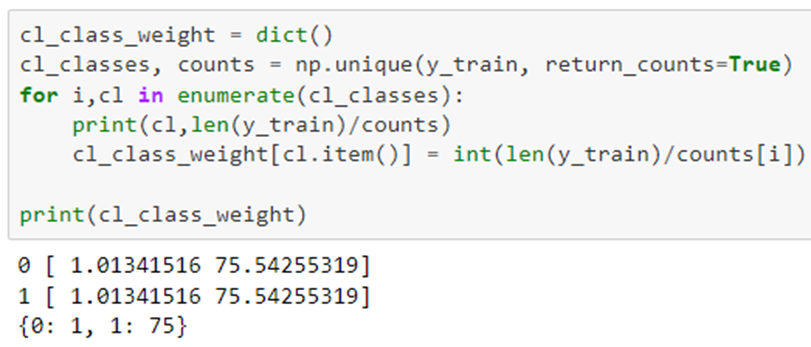}
	\caption{Stage 2 model weights.}
	\label{fig:9}
\end{figure}

The model architecture stack changed slightly. The difference is weighting and a higher dropout at 0.7.  We also improved the stage 2 algorithm for  pre-processing and post-processing. We tested several learning rates, but they were not compatible with our weighting strategy. We experimented with multiple seeds to ensure that the overall results were similar with multiple seeds. Figure~\ref{fig:9} shows the weights used for our model in Stage 2.

\begin{figure}[t!]
	\centering
	\includegraphics[clip, trim=0cm 0cm 0cm 0cm, width=0.8\textwidth]{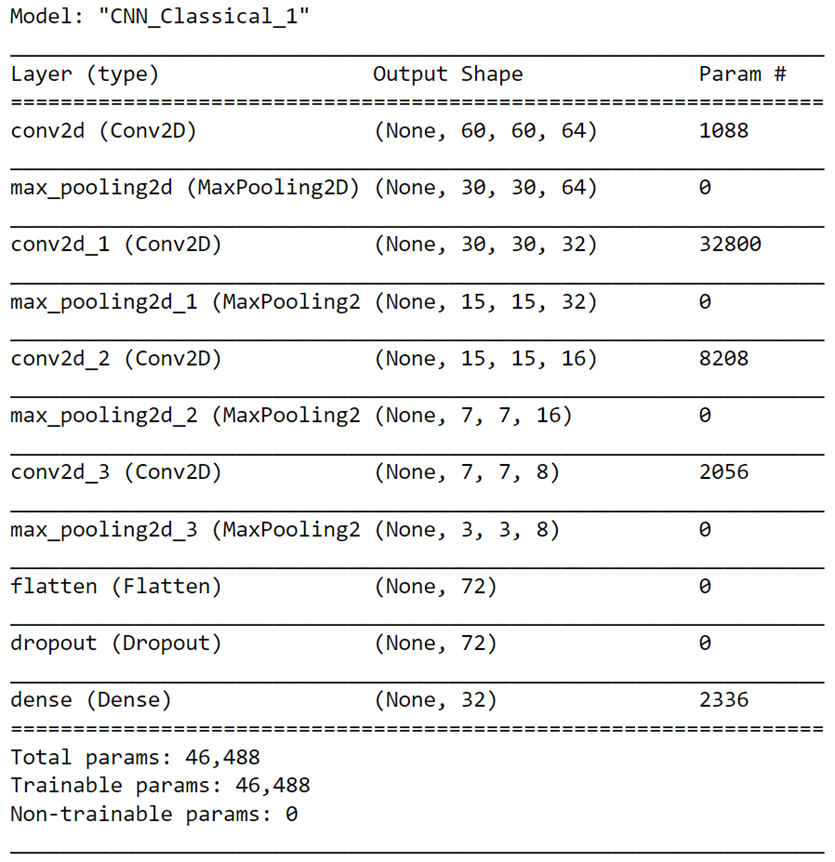}
	\caption{Stage 2 baseline CNN architecture from the KERAS summary.}
	\label{fig:10}
\end{figure}

For both of our experiments we used the similar architecture setups for the classical CNN approach as well as the QNN approach respectively, with only difference being an additional convolutional layer in the CNN model to make up for the quantum convolution. Both setups are shown in the two depictions by Figure~\ref{fig:10} on the example of Experiment 1.

\begin{figure}[t!]
	\centering
	\includegraphics[clip, trim=0cm 0cm 0cm 0cm, width=0.8\textwidth]{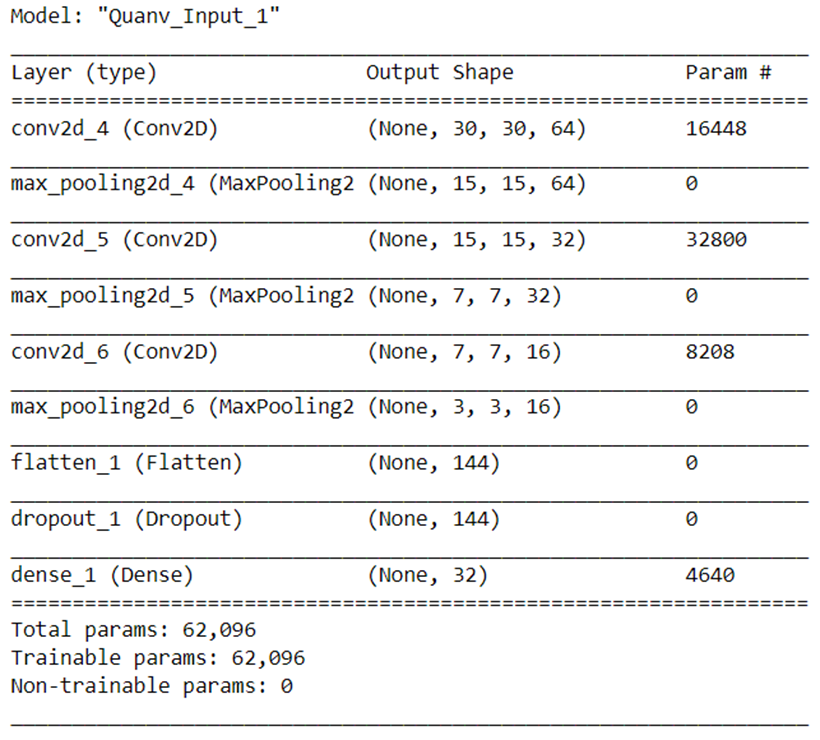}
	\caption{Stage 2 QNN architecture from the KERAS summary.}
	\label{fig:11}
\end{figure}

Figure~\ref{fig:11} illustrates our Stage 2 QNN Model Architecture.

We ran the QNN model on our quantum inputs and the CNN on the normalized images. Each of these first models took 50/50 train/test ratios, while the second iteration of models took 40/60 train/test ratios. We used weighting to up-sample rare classes. Our weighting was not compatible with adjusting learning rates. Therefore, we were able to proceed with the default training rate and let the Adam optimizer do the heavy lifting. We did see some overfitting in the model, so we included a dropout of $0.7$ in our fully connected layers.

\begin{figure}[t!]
	\centering
	\includegraphics[clip, trim=0cm 0cm 0cm 0cm, width=0.7\textwidth]{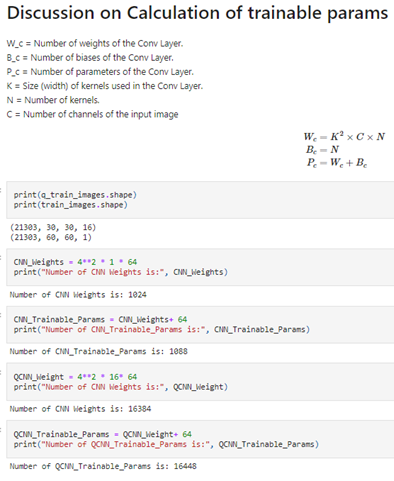}
	\caption{Calculation of trainable parameters in the accompanying JupYter notebook.}
	\label{fig:12}
\end{figure}

We will again include a brief discussion of trainable parameters. The quantum operations are the same as in Stage 1 but the $16$-qubit / image channel output creates a much denser input tensor. It is not necessary to draw the full calculation. Figure~\ref{fig:12} illustrates a computation of Stage 2 trainable parameters and weights.

\section{Empirical Results}
\label{sec:results}
Overall, our results throughout both Studies 1 and 2 indicate that the quantum model can generalize faster, based on fewer input data, and with improved accuracy when compared to the classical model.

\subsection{Results of Stage 1}
The results of Stage 1 show that a quantum model can generalize on a smaller input data set and perform well on a much larger test data set than the classical model can. In a real-world example, this is what we are going for. As stated previously, we tuned our model to produce equal results over $50$ epochs, $100$ training inputs, and 8000 test inputs, and analyzed the results which indicated that the QNN could produce good results with fewer inputs.

This led up to the hypothesis that we can down-sample the concrete data to 50 training images and fit the same model. This action was even more pronounced when you reduce the train inputs to 50 on 8,000 test inputs. The areas inside the boxes in the previous figures showed inconsistency with the CNN model validation. Figure~\ref{fig:res_Stage_1} shows both experiments in the same plot for easier identification:

\begin{figure}[t!]
	\centering
	\includegraphics[clip, trim=0cm 0.7cm 0cm 0cm, width=1\textwidth]{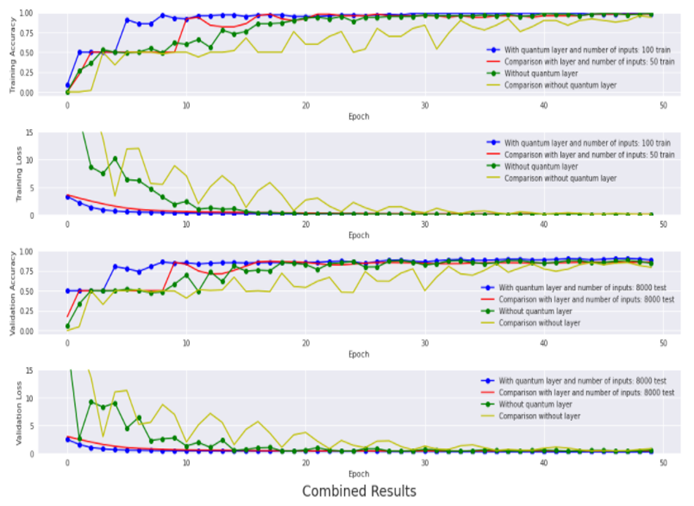}
	\caption{Combined Results of all Experiments in Stage 1}
	\label{fig:res_Stage_1}
\end{figure}

In most real-world examples, it is hard to train a classical model on a small input data set and generalize those results to a larger test data set. This does not seem to be the case in quantum. Our quantum model can train on $40-100$ train images and perform well over $8000$ test images. Under the same conditions, the classical model does not perform nearly as well.

When we started this set of experiments, we did not have any expectations on performance results. However, we gained a lot of evidence to merit further investigation of how QML models can be fit on much less data than classical models can. This means that we have the potential to accurately classify events (such as autonomous driving) using quantum that are not possible with a purely classical CNN approach. Furthermore, this indicates that there is a potential to reduce the need for pre-trained models that are mostly necessary in classical computer vision applications. And lastly, using quantum embeddings, we may achieve levels of performance that have never been possible with classical computing as quantum technology matures.

The breakthrough here is not from a time perspective, or possibly from overall accuracy perspective without performing more experiments in support of our hypothesis. The breakthrough is that a quantum model can optimize after seeing much less data than a classical model can. 

\subsection{Results of Stage 2}
In stage 2, we achieve much higher overall model accuracy with the added benefit of localization in addition to the binary crack classification. Our results also reiterate that quantum models tend to generalize faster, based on fewer input data, and with improved accuracy when compared to classical models.

We conducted two different experiments. The first one was to show how the model performs on an even train/test split, while the second one was executed with a reduced amount of training data to verify our Stage 1 generalization hypotheses with respect to fewer input data.

\subsubsection{Results from Experiment 1: 50/50 Train/Test Split}
Similar to Stage 1, the results indicate that a quantum model can generalize on a smaller input data set and perform well on a much larger test data set than the classical model can, due to the denser input tensor from the quantum embedding. In a real-world example, this is what we are going for. As shown in the chart at the end, we tuned our model to give someone equal results between a classical and quantum example. The blue line shows the quantum while the green shows the classical.

The first two plots show training accuracy/loss. The second two show testing. What we are seeing that is that the accuracy of the quantum model optimizes over fewer epochs, thus seeing the data training data fewer times.

\subsubsection{Results from Experiment 2: 40/60 Train/Test Split}
We ran the same models as the first experiment with fewer images to test our hypothesis that due to the quantum embedding, our denser input tensor can produce an equally or more accurate model with more loss consistency with fewer training images.

The factors that were present in the first experiment where the QNN seems to generalize better and produce a more consistent, accurate model than the CNN were even more pronounced when we down sampled the training data to 50/50. Therefore, our second experiment is in support of our hypothesis that a QNN seems to achieve a similar or greater level of accuracy with more consistent loss than a CNN can while training on fewer inputs. Therefore, further investigation into this topic is deemed valuable.

\begin{figure}[t!]
	\centering
	\includegraphics[clip, trim=0cm 0.7cm 0cm 0cm, width=1\textwidth]{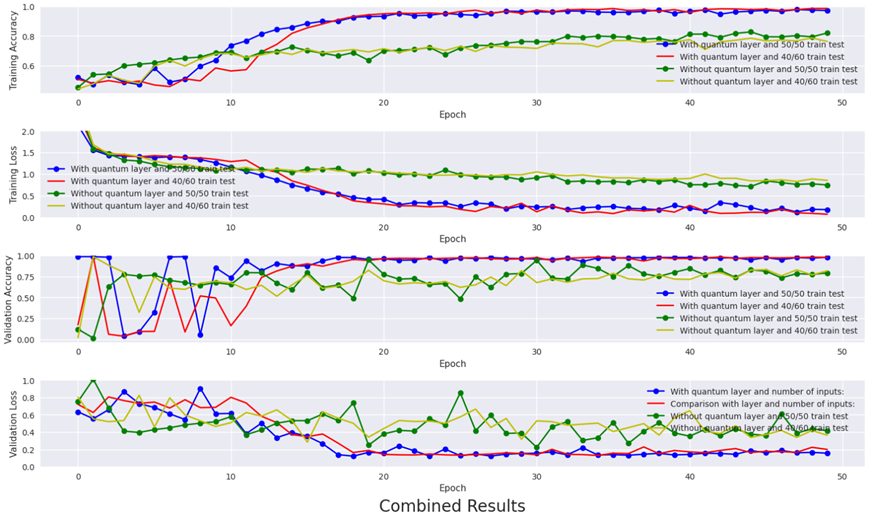}
	\caption{Combined results of all experiments in development stage 2.}
	\label{fig:res_Stage_2}
\end{figure}

Figure~\ref{fig:res_Stage_2} depicts the detailed results and represents our train test splits in Stage 2. The 50/50 split is blue and green for quantum and classical, respectively. The 40/60 split is red and yellow respectively.

\begin{figure}[t!]
	\centering
	\includegraphics[clip, trim=0cm 0cm 0cm 0cm, width=1\textwidth]{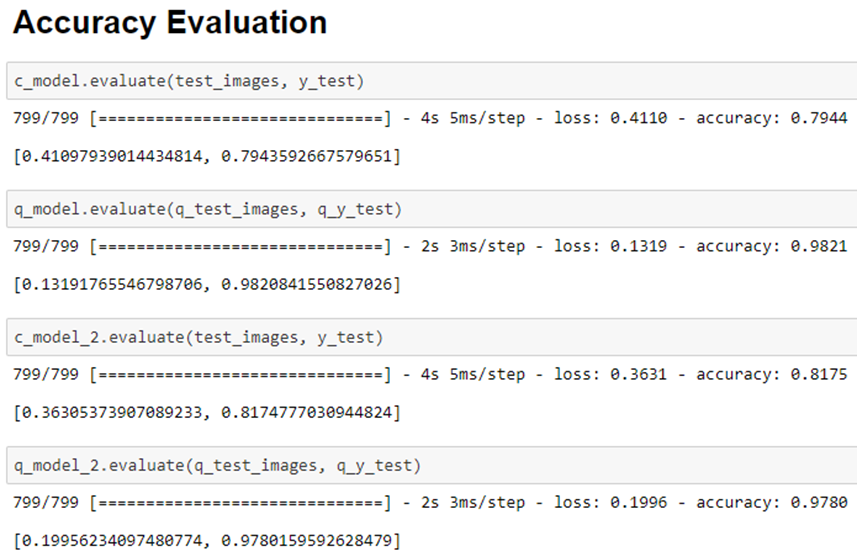}
	\caption{Accuracy from testing in stage 2.}
	\label{fig:res_Stage_2_acc}
\end{figure}

As shown by Figure~\ref{fig:res_Stage_2}, our model produces very accurate outputs on the Stage 2 surface crack image data with a reduced loss value, indicating increased model stability. With our model we achieve a very promising level of 97.8\% validation accuracy on the test data set. The validation accuracy evaluation is shown by Figure~\ref{fig:res_Stage_2_acc}.

\section{Discussion of Results}
\label{sec:discussion}
Our experiments show that the Quanvolutional Neural Network produced more accurate results using less training data. We believe that our approach is a novel and innovative way to localize defects on surface images using quantum technologies via our novel image splitting/labeling algorithm and using our general quantum convolutional approach. This would enable end users to harness the power of quantum on hardware that is currently available.


This means that we have the potential to accurately classify events in a number of cutting-edge image classification use cases using quantum, resolving bottlenecks and significantly improving implementations that have so far relied on purely classical algorithms. Furthermore, this indicates that there is a potential to reduce the effort for pre-training models, e.g., by reducing the number of manually created ground truth masks necessary to label original images, when compared to classical computer vision applications. And lastly, using quantum embeddings, we may achieve levels of performance that have never been possible with classical computing as quantum technology matures.
While we want to do more testing, we have gained additional evidence in Stage 2 over Stage 1 to conclude there are vast benefits from the use quantum technologies in machine learning. Although probably not from a time perspective due to the bottleneck in quantum pre-processing, we have seen evidence to support an overall accuracy perspective with rare events classification and image localization. In future work, we intend to perform more experiments in support of our hypotheses. To conclude, the following highlights summarize our results so far:

\begin{itemize}
    \item We found our image splitting algorithm to be able to provide a localized image classification. We anticipate that with these results, we can use this model on future images, reducing the need to manually label data or create masks.
    \item We were able to detect spots of interest in images in which they were not detected either by human experts nor by existing ML approaches employed by an industrial partner. 
    \item We observed some false negatives produced by our model; these may likely be attributed to the unadjusted weights that we used. However, if we use a less aggressive up-sampling of rare classes we expect to be able to reduce this effect. Alternatively, we could also introduce a regularizer to reduce the number of false negatives. However, false negatives in this case are better than false positives due to the localization. Too many false positives in the CNN localization would deem the entire localization process ineffective, but our model does not have an issue here.
    \item Without the use of quantum, localization is not possible in this way with such a small CNN model architecture.
    \item The QCNN localization model is much more efficient and faster at localization than classical image localization models. This is because the quantum embedding does some of the heavy lifting that a human would otherwise have to perform via manual labeling because the classical CNN cannot generalize too well on such training splits (with respect to our 40/60 experiment).
    \item Overall, we were able to detect almost all of the cracks present (97+\%) and once we adjust the regularizer and/or training weights we anticipate to catch all of them (99+\%).
\end{itemize}

\section{Conclusion and Outlook}
\label{sec:conclusion}
The model properties we have used to compute the effective dimension, detailed within the references provided, can also be used in other ways to aspects of quantum neural networks. There exist other potential research and development opportunities: First the evaluation of the effect of model regularization on capacity and trainability measures, second the analysis of trade-offs between capacity and generalizability during training to assist hyperparameter optimization, and third the extension of capacity measures to variational quantum neural networks.

The theoretical methods and tools we have developed to analyze capacity must be considered in combination with other factors such as cost, expected quantum computing development timelines and other ML measures to assess when QA and other processes could or should be integrated into a full end-to-end QML process.



\bibliographystyle{unsrt}
\bibliography{main}

\begin{thebibliography}{10}

\bibitem{Oyallon2019-SNF}
E.~Oyallon, S.~Zagoruyko, G.~Huang, N.~Komodakis, S.~Lacoste-Julien,
  M.~Blaschko, and E.~Belilovsky.
\newblock {Scattering Networks for Hybrid Representation Learning}.
\newblock {\em IEEE Trans. on Pattern Analysis \& Machine Intelligence}, 41(9),
  2019.

\bibitem{Karniadakis2021-PIM}
G.E. Karniadakis, I.G. Kevrekidis, L.~Lu, P.~Perdikaris, S.~Wang, and L.~Yang.
\newblock {Physics-Informed Machine Learning}.
\newblock {\em Nature Reviews Physics}, 3, 2021.

\bibitem{vRueden2021-IML}
L.~von Rueden~{et al.}
\newblock {Informed Machine Learning - A Taxonomy and Survey of Integrating
  Prior Knowledge into Learning Systems}.
\newblock {\em IEEE Trans. on Knowledge and Data Engineering}, 2021.

\bibitem{Dunjko2016quantum}
V.~Dunjko, J.M. Taylor, and H.J. Briegel.
\newblock {Quantum-Enhanced Machine Learning}.
\newblock {\em Physical Review Letters}, 117(13), 2016.

\bibitem{Kerenidis_2019}
I.~Kerenidis, J.~Landman, and A.~Prakash.
\newblock {Quantum Algorithms for Deep Convolutional Neural Networks}.
\newblock {\em arXiv:1911.01117 [quant-ph]}, 2019.

\bibitem{Lue_2021}
Y.~L{\"u}, Q.~Gao, J.~L{\"u}, M.~Ogorzalek, and J.~Zheng.
\newblock {A Quantum Convolutional Neural Network for Image Classification}.
\newblock {\em arXiv:2107.03630v2 [quant-ph]}, 2021.

\bibitem{Henderson2021-QNN}
M.~Henderson, S.~Shakya, S.and~Pradhan, and T.~Cook.
\newblock {Quanvolutional Neural Networks: Powering Image Recognition with
  Quantum Circuits}.
\newblock {\em Quantum Machine Intelligence}, 2, 2021.

\bibitem{Henderson2021-MFA}
M.~Henderson, J.~Gallina, and M.~Brett.
\newblock {Methods for Accelerating Geospatial Data Processing Using Quantum
  Computers}.
\newblock {\em Quantum Machine Intelligence}, 3, 2021.

\bibitem{Mathur2021-MIC}
N.~Mathur, J.~Landman, Y.Y. Li, and M.~Strahm.
\newblock {Medical Image Classification via Quantum Neural Networks}.
\newblock {\em arXiv:2109.01831 [quant-ph]}, 2021.

\bibitem{Mattern_2021}
D.~Mattern, D.~Martyniuk, H.~Willems, F.~Bergmann, and A.~Paschke.
\newblock {Variational Quanvolutional Neural Networks with Enhanced Image
  Encoding}.
\newblock {\em arXiv:2106.07327 [cs.CV]}, 2021.

\bibitem{Cong_etal_2019}
I.~Cong, S.~Choi, and M.D. Lukin.
\newblock {Quantum Convolutional Neural Networks}.
\newblock {\em arXiv:1810.03787 [quant-ph]}, 2019.

\bibitem{Preskill_2018}
J.~Preskill.
\newblock {Quantum Computing in the {NISQ} Era and Beyond}.
\newblock {\em Quantum}, 2, 2018.

\bibitem{KingmaBa}
D.~P. Kingma and J.~Ba.
\newblock {Adam: A Method for Stochastic Optimization}.
\newblock In {\em Proc. ICLR}, 2015.

\end{thebibliography}

\end{document}